\documentclass[aps]{revtex4}

\begin{document}

\def\bra{\langle}  \def\ket{\rangle}

\title{Anomalous quantum states in 
finite macroscopic systems
}

\author{Akira Shimizu}
\address{Department of Basic Science, University of Tokyo,
3-8-1 Komaba, Tokyo 153-8902, Japan\\
and
PRESTO, JST, 4-1-8 Honcho Kawaguchi, Saitama, Japan\\
shmz@ASone.c.u-tokyo.ac.jp}

\author{Takayuki Miyadera}
\address{Department of Information Sciences, Tokyo University of Science,
Chiba 278-8510, Japan\\
miyadera@is.noda.tus.ac.jp}  

\author{Akihisa Ukena}
\address{Department of Basic Science, University of Tokyo,
3-8-1 Komaba, Tokyo 153-8902, Japan\\
ukena@ASone.c.u-tokyo.ac.jp}

\begin{abstract}
We consider finite macroscopic systems,
i.e., systems of large but finite degrees of freedom,
which we believe are poorly understood as compared
with small systems and infinite systems.
We focus on pure states 
that do not have the `cluster property.'
Such a pure state is entangled {\em macroscopically}, and 
is quite anomalous in view of many-body physics
because it does not approach any pure state in the 
infinite-size limit.
However, we often encounter such anomalous states
when studying finite macroscopic systems, 
such as quantum computers with many qubits and
finite systems that will exhibit symmetry breaking in 
the infinite-size limit.
We study stabilities of such anomalous states
in general systems.
In contrast to the previous works, 
we obtain general and universal results, 
by making full use of the locality of the theory.
Using the general results, we discuss roles of anomalous 
states in quantum computers, and 
the mechanism of emergence of a classical world from quantum theory.
\end{abstract}

\maketitle

\section{Introduction}

In quantum information theory, one is usually interested in 
the asymptotic behaviors of quantities of interest as 
the size $N$ of the data is increased.
To study the asymptotic behavior, one must take the data size 
large but finite.
Therefore, 
quantum information theory treats quantum 
systems that have large but {\em finite} degrees of freedom, 
$1 \ll N < +\infty$.
In quantum computers, in particular, the large degrees of freedom 
extend spatially over regions whose sizes $\propto N$.
In quantum-information-theoretical terms, such a system 
may be called a {\it many-partite system}, whereas 
in many-body-theoretical terms it is called a 
{\it finite macroscopic system}.

Finite macroscopic systems are poorly understood as compared
with small systems and infinite systems.
In particular, there exist anomalous pure states
that do not exist in infinite systems.
Namely, in finite systems there exist pure states 
that do not approach any pure states of infinite systems
as $N \to \infty$ \cite{SM02}.
It is interesting to explore 
properties of such anomalous states
as well as their roles in quantum information theory.

In this short review, we point out that
such states are {\em macroscopically} entangled.
Two measures of the macroscopic entanglement are described.
They are shown to be related directly to stabilities of 
quantum states.
These general results are applied to analyses of 
quantum computers and of the mechanism of 
emergence of thermodynamical world from quantum theory.

Throughout this review, we assume that the number of qubits $N$ is 
large but finite. Hence, theories of infinite systems are not directly 
applicable.
Since we focus on asymptotic behaviors of physical properties 
for huge $N$, prefactors are not important.
Furthermore, only pure states are considered, although 
the result of section \ref{ss-lm} is applicable to mixed states as well.

\section{Proposed measures of macroscopic entanglement}

In an previous paper \cite{SM02}, we 
proposed two measures of macroscopic entanglement, 
which are closely related to each other, 
for quantum states in finite macroscopic systems, or, equivalently, 
in many-partite systems.
We here summarize them and explain their physical meanings.

\subsection{Cluster property}

The first measure 
is the `cluster property' in finite systems.
We recall that in {\em infinite} systems it
is defined as follows \cite{haag}:
A quantum state is said to have the cluster property
if 
\begin{equation}
\langle \delta \hat a(x) \delta \hat b(y) \rangle
\to 0
\mbox{ as } |x - y| \to \infty
\end{equation}
for {\em any} local operators $\hat a(x)$ and $\hat b(y)$
at positions $x$ and $y$, respectively, 
where 
$
\delta \hat a(x) \equiv \hat a(x) - \langle \hat a(x) \rangle
$
and 
$
\delta \hat b(y) \equiv \hat b(y) - \langle \hat b(y) \rangle
$ \cite{confuse}.
%
Here, by an {\em local operator at $x$} 
we mean a operator on a local Hilbert space at $x$ \cite{cp}.

Since we are interested in finite macroscopic systems, for which 
$1 \ll N < +\infty$, 
we generalized the concept of the cluster property to
the case of finite systems.
It reads roughly as follows:
(The precise definition is described in Ref.\ \cite{SM02}.)
Let $\Omega(\epsilon)$ be the size of the region
outside which correlations of fluctuations of 
any local operators become smaller 
than a small positive number $\epsilon$.
We consider a sequence of systems
with various values of $N$ and associated states, 
where the states with different values of $N$ are similar to each other
\cite{similar}.
We say that the states (for large $N$) of the sequence
have the cluster property 
if for any $\epsilon > 0$ one can make 
$\Omega(\epsilon)$ to be independent of $N$
by taking $N$ large enough.

The lack of the cluster property means the existence of 
a long-range correlation(s).
Therefore, for a pure state, it means an entanglement.
Since a small number of Bell pairs do not destroy 
the cluster property (see below), 
the lack of the cluster property
means a {\em macroscopic} entanglement.
Such a macroscopically entangled state (i.e., 
a {\em pure} state without the cluster property)
is quite anomalous in the sense that it does {\em not} approach 
any pure state as $N \to \infty$.
This can be seen from a fundamental theorem \cite{haag}:
For {\em infinite} quantum systems, 
any pure state has the cluster property.

\subsection{Fluctuations of additive operators}

The second, probably more practical, 
measure of macroscopic entanglement which we
proposed in Ref.~\cite{SM02} is
fluctuations of additive operators.
An operator $\hat A$ is called additive if
it is the sum of local operators;
\begin{equation}
\hat A = \sum_x \hat a(x).
\end{equation}
Here, $\hat a(x')$ ($x' \neq x$) is not necessarily a spatially translated 
one of $\hat a(x)$:
e.g., $\hat a(x) = \hat \sigma_y(x)$ whereas 
$\hat a(x') = \hat \sigma_z(x')$.
We note that the fluctuation of $\hat A$ is the sum of two-point correlations;
\begin{equation}
\bra \psi | (\Delta \hat A)^2 | \psi \ket
=
\sum_x \sum_y
\bra \psi | \Delta \hat a(x) \Delta \hat a(y) | \psi \ket,
\end{equation}
whose value depends on the choices of $\hat a(x)$ and 
of the state $| \psi \ket$.
Assuming that $\hat a(x)$ is bounded, 
we normalize $\hat a(x)$ as 
$\| \hat a(x) \| = 1$ \cite{norm}.
For a given $| \psi \ket$, 
we consider the supreme value
\[ 
\sup_{\{\hat a(x)\}} 
\bra \psi | (\Delta \hat A)^2 | \psi \ket
\]
under this normalization condition.
Since it is proportional to the number of 
correlated pairs of points in $| \psi \ket$, 
and since we are interested in asymptotic behaviors for large $N$, 
we define an index $p$ by
\begin{equation}
\sup_{\{\hat a(x)\}}
 \bra \psi | (\Delta \hat A)^2 | \psi \ket
= O(N^p).
\end{equation}
If $p=1$ we call $| \psi \ket$ a normally-fluctuating state (NFS), 
whereas if $p$ takes the maximum value $p=2$ we call $| \psi \ket$
an anomalously-fluctuating state (AFS).
It can be shown that a NFS 
has the cluster property, whereas 
an AFS does not.
Therefore, AFSs are quite anomalous states which are
entangled macroscopically.

We can easily show that 
$p=1$ for any separable states. Hence, all separable states
are NFSs. 
Note however that the inverse is not necessarily true.
For example, 
a many-body state with a single Bell pair,
$\displaystyle \frac{1}{\sqrt{2}} \left( 
|100 \cdots 00 \ket 
+ |000\cdots 01 \ket
\right)$, 
and the `W state,' 
\begin{equation}
| {\rm W} \rangle
\equiv 
\frac{1}{\sqrt{N}}
\left[
|100 \cdots 0 \ket
+|010 \cdots 0 \ket
+|001 \cdots 0 \ket
+ \cdots
+|000 \cdots 1 \ket
\right],
\end{equation}
are non-separable, but are NFSs having the cluster property because
$\displaystyle 
\sup_{\{\hat a(x)\}} \bra \psi | (\Delta \hat A)^2 | \psi \ket
= \mbox{constant} \times N + O(1)
$ 
hence $p=1$.
These states {\em are} entangled, but {\em not} macroscopically.

`Vacuum states' with a broken continuous symmetry
are generally more entangled, for which $1 < p < 2$ due to fluctuations 
through Nambu-Goldstone modes.

The simplest example of an AFS ($p=2$) is 
a `cat state,' $\displaystyle | {\rm C} \ket = \frac{1}{\sqrt{2}} \left( 
|000 \cdots 0 \ket + |111\cdots 1 \ket
\right)$.
This state is macroscopically entangled:
In fact, $| {\rm C} \ket$ violates Bell's inequality by 
a macroscopic factor $2^{(N-1)/2}$ \cite{mermin}.
Although the macroscopic entanglement of
this state may be obvious
(because it has a very simple form), 
we can investigate 
whether more complicated pure states are macroscopically entangled
by calculating $p$.

We can thus use the cluster property and the value of $p$ 
as measures of macroscopic entanglement
for pure states.

\subsection{Advantages of the proposed measures}

As compared with other measures or definitions 
of many-partite entanglement, 
the proposed measures of macroscopic entanglement
are related most directly to many-body physics.
Since many-body physics has been developed along with many experiments, 
the proposed measures are also related 
directly to properties of real systems.
For example, 
some other definitions of entanglement 
classify the state $|{\rm W} \ket$ 
as more entangled than $| {\rm C} \ket$.
However, $|{\rm W} \ket$ is quite a normal state in many-body physics 
and experiments:
It represents, e.g., a low-lying excited state of an insulating solid, 
in which a single Frenkel exciton is excited on the ground state.
Such a state can be easily generated experimentally.
It is therefore more reasonable to classify $| {\rm W} \rangle$
as a normal state whose entanglement is negligible for 
macroscopic properties.
This is in accordance with the proposed measures:
$|{\rm W} \ket$ is an NFS ($p=1$) having the cluster property, hence 
is {\em not} entangled macroscopically.

Furthermore, 
an efficient method of computing $p$ for general states 
has been developed in Ref.~\cite{SS03}.
In contrast, some other measures of entanglement are 
hard to compute for general systems.
On the other hand, a disadvantage of the present measures is that 
they can be used for pure states only.

\section{Physical properties of macroscopically entangled states}

When studying finite macroscopic systems, we often encounter AFSs.
For example, consider a finite system which will exhibit
symmetry breaking as $N \to \infty$.
It is known that 
when the order parameter does not commute with the Hamiltonian of 
the system the {\em exact} ground state 
for finite $N$ 
is not 
a symmetry-breaking state but the {\em symmetric} ground state that 
possesses all the symmetries of the Hamiltonian \cite{HL,KT,pre01}.
The well-known diagram, in which symmetry-breaking states have 
the lowest energy, is a misleading result of a mean-field approximation.
Anomalous states, such as AFSs, cannot be obtained by 
the mean-field approximation.
AFSs appear also in quantum computers \cite{us},
as will be described briefly in section \ref{ss-qc}.
Therefore, it is important to explore 
properties of anomalous states, such as AFSs, 
of finite macroscopic systems.

For this purpose, we studied 
the stability of quantum states of 
general macroscopic systems of
finite sizes, against weak classical noises, 
weak perturbations from environments, 
and local measurements \cite{SM02}.
In contrast to the previous works on similar subjects, 
we obtained general and universal results which are 
independent of details of models, 
by making full use of the locality of the theory.

\subsection{Decoherence by noises or environments}

We say that a pure state is `fragile' if its decoherence rate 
$\Gamma$ is 
anomalously great in the sense that 
\begin{equation}
\Gamma \sim 
K N^{1+\delta},
\label{KN1pd}\end{equation}
where $K$ is a function of microscopic parameters,
and $\delta$ is a positive constant.
This is an anomalous situation in which 
the decoherence rate {\em per qubit}, 
$
\Gamma/N
$,
grows with increasing $N$ as $ \sim K N^{\delta}$.
In contrast, $\delta = 0$ is a normal situation.
We showed that 
NFSs never become fragile in weak perturbations from 
{\em any} noises or environments \cite{SM02}.
Hence, the non-fragility of NFSs, 
which are typical normal states \cite{normal}, 
against decoherence was shown most generally, independently of
details of models,  for the first time.

Regarding fragility of AFSs, 
on the other hand, 
we found that 
an AFS becomes {\em either fragile or non-fragile} 
depending on the spectral intensities of 
the noises or correlation functions in the environments \cite{SM02}.
Although one might expect that anomalous states such as 
AFSs would always be fragile, 
our general result is against such naive expectations.
This indicates that we must 
go beyond the decoherence rate
in order to study the stability of anomalous states.
For this purpose, 
we proposed the new criterion of stability, which will be explained 
in the following section.

\subsection{Stability against local measurements}
\label{ss-lm}

We say a quantum state is 
`stable against local measurements' if
the result of any local measurement is not affected by 
any local measurement at a distant point.
(The precise definition is described in Ref.\ \cite{SM02}.)
According to experiences, macroscopic states 
seem to have this stability, i.e., 
they do not change significantly by 
measurement of only a tiny part of a macroscopic system.
Indeed, we successfully proved the following theorem \cite{SM02}:
{\em If a quantum state has the cluster property,
then it is stable against local measurements, and vice versa.}

Namely, changes induced by any 
local measurements are small for normal states 
that have the cluster property.
In contrast, 
anomalous states that do not have 
the cluster property, such as AFSs, are changed drastically 
by measurements of some relevant local observables.
Such drastic changes continue by 
repeating measurements of relevant local observables, until
the state becomes a state with the cluster property.
We conjectured (and confirmed in several examples) that
the number of local measurements necessary for reducing an AFS to 
a state with the cluster property would be much less than $N$.

\section{Applications}

We have successfully clarified 
what stabilities are related to what anomalies of 
quantum states of general systems.
Using these general results, we can discuss 
the origin of symmetry-breaking in finite systems \cite{SM02},
properties of quantum chaotic systems with 
large degrees of freedom \cite{SS03}, and so on.
Among such many applications, we here describe 
roles of anomalous states in quantum computers, 
and the mechanism of emergence of a classical macroscopic 
world from quantum theory.

\subsection{Application to quantum computers}
\label{ss-qc}

In Ref.~\cite{us}, we analyzed 
quantum computers which perform Shor's factoring 
algorithm, paying attention to asymptotic properties as 
the number $L$ of qubits is increased.
Using numerical simulations and the general theory described 
above, we showed the following:
AFSs appear in various stages of 
the computation. For large $L$, they decohere at anomalously 
great rates by weak noises that simulate some noises in real systems.
Decoherence of some of the AFSs is fatal to the results of the computation, 
whereas decoherence of another some of the AFSs does not have
strong influence on the results of the computation.

Therefore, there are 
three classes of states in Shor's factoring algorithm:
(i) NFSs, such as the state after the Hadamard transformation
$|\psi_{\rm HT} \ket$, 
for which 
the reduction of fidelity $F$ behaves as $1-F \propto$ $L$, 
and 
the reduction of the success probability $T$ behaves as 
$T_{\rm clean} -T \propto$ $L$.
Here, the success probability $T$ means the probability 
of getting a successful result by a single computation, 
and $T_{\rm clean}$ denotes its value when the noises are absent \cite{us}.
(ii) Non-crucial AFSs, such as the final state (before the measurement
to obtain a computational result)
$|\psi_{\rm final} \ket$, 
for which $1 -F \propto$ $L^{2}$ 
whereas $T_{\rm clean} -T \propto L$.
(iii) Crucial AFSs, such as the state after the modular 
exponentiation $|\psi_{\rm ME} \ket$, 
for which 
$1-F \propto$ $L^{2}$ and
$T_{\rm clean} -T \propto L^{2}$.

We consider that the exponential speedup over classical computers
is achieved by the use of macroscopic entanglement of the crucial AFSs.

When a crucial AFS decoheres, 
the probability of getting a correct computational result is 
reduced in approximately proportional to $L^2$.
The reduction thus becomes anomalously large
with increasing $L$, even when the coupling constant to the noise
is rather small.
Since error correcting codes and decoherence-free subspaces
are not almighty in real physical systems, 
it is desirable that quantum computers 
should be improved in such a way that
all AFSs appearing in the algorithms do not 
decohere at such great rates in the presence of realistic noises.

\subsection{Emergence of a classical world}

Thermodynamics assumes that 
$\bra (\Delta A)^2 \ket = o(N^2)$
for all additive quantities, in any pure thermodynamical phase.
(Otherwise, predictions about averages $\bra A \ket$ would 
have no meaning.)
In quantum theory, however, 
there exist pure states, which we call AFSs,
that do not satisfy this assumption because by definition 
$\bra (\Delta \hat A)^2 \ket = O(N^2)$
for some additive operator(s).
Hence, questions arise: 
Why is thermodynamics applicable so universally to macroscopic 
systems?
How a thermodynamical world, in which 
$\bra (\Delta A)^2 \ket = o(N^2)$ for all additive quantities, 
emerges from quantum theory?
We can answer these questions using the theorem of section \ref{ss-lm}.

Suppose that an AFS happens to be realized in a laboratory.
An experimentalist will be pleased to have such an anomalous 
state, and he will decide to perform experiments on it.
He will cautiously start with local measurements on 
some tiny parts of the system to obtain 
some preliminary informations on the state, expecting that 
such local perturbations would not alter the state of the 
macroscopic system significantly.
But, according to the theorem of section \ref{ss-lm}, 
measurements even on tiny parts of the system will 
alter the state drastically if the state is 
an AFS.
As a result, the state after not so many ($\ll N$) local 
measurements would become a state with the cluster property, 
for which $\bra (\Delta \hat A)^2 \ket = o(N^2)$
in consistency with thermodynamics.
Thus, a thermodynamical world almost always emerges.

%


\begin{thebibliography}{99}


\bibitem{SM02}
A.\ Shimizu and T.\ Miyadera, 
Phys.\ Rev.\ Lett.\ {\bf 89}, 270403 (2002). 


\bibitem{haag}
R.\ Haag, {\it Local Quantum Physics} (Springer, Berlin, 1992).

\bibitem{confuse}
The cluster property should not be confused with
the absence of the long-range order.
For example, symmetry-breaking vacua 
have {\em both} the long-range order and 
the cluster property \cite{SM02,haag}.

\bibitem{cp}
There is another way of defining local operators, hence another 
way of defining the cluster property:
A local operator {\em around} $x$ is an operator 
on local Hilbert spaces
at positions
in any finite region around $x$.
A quantum state is said to have the cluster property
if 
$ 
\langle \delta \hat a(x) \delta \hat b(y) \rangle
\to 0
\mbox{ as } |x - y| \to \infty
$ 
for any local operators $\hat a(x)$ and $\hat b(y)$
{\em around} $x$ and $y$, respectively.
Ref.\ \cite{haag} employed this definition.
If a state has the cluster property in this sense, 
then it also has the cluster property defined in the text.


\bibitem{similar}
For example, consider quantum computers of various sizes which all
perform the same algorithm with different sizes of inputs.
We demonstrated in Ref.\ \cite{us} that in this case 
a state in a computer of some $N$ is similar to 
a state(s) in a computer of another $N$, 
because they are generated by the same algorithm.
%
Another example is the ground-state wavefunctions 
of many particles in a sphere for the same particle density, 
for various sizes of spheres.

\bibitem{norm}
$\| \hat a \|$ denotes the norm 
of an operator $\hat a$, which is defined by
$ 
\| \hat a \|
\equiv
\sup\{ \| \hat a | \phi \ket \|; \bra \phi | \phi \ket =1 \}
$. 

\bibitem{SS03}
A.\ Sugita and A.\ Shimizu, 
e-print: quant-ph/0309217.

\bibitem{mermin}
N. D. Mermin, 
Phys. Rev. Lett. {\bf 65}, 1838 (1990).


\bibitem{HL}
P.\ Horsh and W.\ von der Linden, Z.\ Phys.\ {\bf B72}, 181 (1988).

\bibitem{KT}
T. Koma and H. Tasaki, 
J.\ Stat.\ Phys.\ {\bf 76}, 745 (1994).

\bibitem{pre01}
A.\ Shimizu\ and\ T.\ Miyadera, 
Phys.\ Rev.\ {\bf E64}, 056121 (2001).

\bibitem{us}
A.\ Ukena and A.\ Shimizu, to appear in Phys. Rev. A; 
e-print quant-ph/0308005.

\bibitem{normal}
Here, the `normal state' does not mean the normal state in 
the physics of superconductors:
The BCS state is a normal state in the sense of the present paper.

\end{thebibliography}
\end{document}